\documentclass[aps,prb,showkeys,showpacs]{revtex4}
\usepackage{amsmath}%
\usepackage{amsfonts}%
\usepackage{amssymb}%
\begin{document}
\title{Generalized Slater-Jastrow trial functional: Application to high order correlations in the two and three dimensional electron gas}
\author{James C. Porter}
\affiliation{Department of Physics and Astronomy, Eastern Michigan University,\\ Ypsilanti, MI 48197 \email {jporter@emich.edu}}
\begin{abstract}
The GSJ trial functional is a modification of the Slater-Jastrow functional where, effectively, the argument of the Jastrow factor can be momentum dependent. The associated Euler equations, which provide an alternative to the usual approach via perturbation theory, are solved to yield the first three terms of the electron correlation energies for the two and three dimensional electron gas at high density. In two dimensions we find $\epsilon_c=-0.4039-0.50\,r_s\,\log r_s$ plus an estimated $-0.87\,r_s$, and for three dimensions $\epsilon_c=0.0622\,\log\,r_s-0.082+0.059\,r_s\,\log r_s$ plus an estimated $-0.015\,r_s$ where $r_s$ is the applicable Seitz radius in each case. The present paper extends the findings of the similarly titled 2009 work (Phys. Rev. B 80)  in two respects. First the two dimensional as well as the three dimensional electron gas is treated and second high order coulomb correlations are obtained in each case, furnishing an alternative to the findings of Gell-Mann and Breuckner (Phys. Rev. 106 (1957)), Carr and Maradudin (Phys. Rev. 133 (1964)) and Rajagopal and Kimball (Phys. Rev. B 15 (1977)).
The latter development is due to the discovery of an analytic solution for the lowest order two electron functions, obtained previously only numerically, and a procedure which isolates the logarithmic divergence in the three dimensional case without the need to set the Seitz radius to zero. 
 Other points of interest include a more complete accounting of background energy cancelation in the second section, and the evaluation of a 12-dimensional integral contributing to the high order coulomb correlation energy in Appendix B.
\end{abstract}
\maketitle
\subsection*{Introduction}
Variational approximations to the wave function satisfying the many body Schrodinger equation can be obtained from a great variety of comparison functionals\cite{rev}.  A  subset of these, consists of a series which begins with the Hartree Fock (HF) functional, an antisymmetrized product of $N$ one electron functions often written as a determinant,\cite{Slater} and continues with the GSJ functional,\cite{P09} constructed from both one and two electron functions. The direct terms of the functionals in this series approximate the many electron wave function in a manner suggested by the Mayer cluster expansion,\cite{M} in that the Hartree two electron density matrix is approximated as the product of two one electron matrices while the direct terms of the GSJ functional approximate the three electron density matrix by terms representing the three ways of combining one and two electron matrices less twice the product of three one electron matrices.\cite{P65} 

Like the one-electron functions of the HF functional, the one and two-electron functions of the GSJ functional are a priory unconstrained, save for simple orthogonality and normalization requirements, and are determined by associated integro-differential Euler equations. Also, like the HF functional, the coordinates (position and momentum) of a given electron appear only once in each additive term of the GSJ functional.

The Euler equations associated with the GSJ functional are tractable in the case of the homogeneous electron gas in a background of compensating positive charge (possibly with the addition of an externally imposed electric field.\cite{ORD}) Here plane waves satisfy the one electron Euler equation, as in the HF theory,\cite{FW} and the two electron Euler equation has an approximate solution in closed form which allows evaluation of the order $r_s\,\log\,r_s$ term of the correlation energy, $r_s$ being the applicable Seitz radius. It is instructive to compare these results with those of perturbation theory. The latter assumes that the system energy is expressible as a power series in some small parameter measuring the ratio of potential to kinetic energies. This assumption does not hold in the case of the electron gas because of the energies logarithmic dependence on the assumed small Seitz radius so that complicated\cite{GB,Carr} and possibly problematic\cite{Dirac} work-arounds are required.

The following sections cover the construction of the GSJ functional, Euler equations for the electron gas Hamiltonian, their solution and correlation energy calculations through the $r_s\, \log\,r_s$ term, with an estimate for the order $r_s$ term. Results for both the two and three dimensional gas are presented in parallel prior to the evaluation of the individual correlation energies.

\subsection*{GSJ functional and energy integral}

Denoting the one and two electron functions by $U(j_1;i_1)$ and  $f(j_1,j_2;i_1,i_2)$, where indices $j_1,\,j_2 \cdots$ appearing prior to the semicolons label momenta and spin, and integers $i_1,\,i_2\cdots$ following the semicolons show position dependences $\mathbf{r}_{i_1},\,\mathbf{r}_{i_2}\cdots$, the GSJ functional is constructed from the Laplace development\cite{KK} of the N electron HF determinant, $\Psi_{HF}=\det((U(j;i)))$, by 2n-rowed minors,
\begin{equation}
\Psi_{HF}=\sum_{1\leq j_1<j_2<\cdots j_{2n}\leq N}M(j_1,j_2,\cdots j_{2n};i_1,i_2\cdots i_{2n})D(j_1,j_2,\cdots j_{2n};i_1,i_2\cdots i_{2n}),\label{eq.1}
\end{equation}
where $M$ is the minor obtained by retaining only those elements of $\Psi_{HF}$ found in columns $j_1,j_2.\cdots j_{2n}$ and rows $i_1,i_2,\cdots i_{2n}$ and where the $N\!\!-\!2n$ rowed determinant $D(j_1,j_2,\cdots j_{2n};i_1,i_2\cdots i_{2n})$ is the corresponding cofactor. 

$\Psi_{GSJ}$ is obtained from Eq.(1) upon replacing the minors there by $1/n!$ times antisymmetrized products of $n$ two electron functions and summing over all distinguishable combinations of the row numbers chosen for the Laplace developments. A final summation over $n=0,1,\,2,\cdots [(N-1)/2]$, where the $n=0$ term is just the HF determinant itself, gives the GSJ series\cite{thesis,note}
\begin{eqnarray}  
\Psi_{GSJ}=\sum_{n=0}^{[(N-1)/2]}\frac{1}{n!}\sum_{\stackrel{\scriptstyle 
1 \leq j_{1} < j_{2} < \cdots <j_{2n} \leq N}{1 \leq i_{1}  <  i_{2} <\cdots <i_{2n}\leq N}} (\hat{A}^j\hat{A}_i\prod_{k=1}^n f(j_{2k-1},j_{2k};i_{2k-1},i_{2k})\frac{}{})\nonumber \\
\times D(j_1,\cdots,j_{2n};i_1,\cdots,i_{2n}),\qquad\label{eq.2}
\end{eqnarray}
where $\hat{A}^j$ and  $\hat{A}_i$ are commuting antisymmetrizing operators acting on momenta and positions respectively.

The Coulomb correlation energy is given by $1/N$ times the difference between $E_{GSJ}$, the energy integral, and the corresponding HF estimate, where 
\begin{eqnarray}
E_{GSJ}=(\Psi_{GSJ},\left[-\frac{\hbar^2}{2m}\sum_{1\leq i\leq N}\mathbf{\nabla}_{i}^2+\frac{1}{2}\sum_{1\leq i_1\neq i_2\leq N}\phi(i_1,i_2)\right]\Psi_{GSJ})/(\Psi_{GSJ},\Psi_{GSJ})\nonumber \\
+E_{background}.\label{eq.3}
\end{eqnarray}

As in Slater's derivation of the HF equations,\cite{Slater} the evaluation of $E_{GSJ}$ is facilitated by the use of Lagrange multipliers $\{\epsilon(j)\}$ to constrain the one and two electron functions.  The constraints imposed are that the one electron functions for an orthonormal set and that
\begin{subequations}
\begin{equation}
\int d^Dr_1 d^D r_2 U^{\dagger}(j_1;1)U^{\dagger}(j_2;2)\hat{A}^j\hat{A}_if(j_1,j_2;1,2)=0, \label{eq.4a}
\end{equation}
where the spatial integrals may be over either two or three dimensions, $D=2$ or $D=3$. Since $f(j_1,j_2;1,2)\equiv f(j_2,j_1;2,1)$, Eq.(3a) can be written more simply in terms of $F(j_1,j_2;1,2)\equiv 2f(j_1,j_2;1,2)$ as
\begin{equation}
\int d^Dr_1 d^D r_2 U^{\dagger}(j_1;1)U^{\dagger}(j_2;2)\hat{A}^jF(j_1,j_2;1,2)=0. \label{eq.4b}
\end{equation}
\end{subequations}

Terms, where $n$ in Eq.(2) is much less than $N$ but also much greater than unity, dominate in Eq.(3), where the large numeric factors which appear in both numerator and denominator reach broad maxima. It can be shown\cite{P65, thesis} that the quotient of these multipliers can be dropped, and the GSJ series replaced by just its first two terms, \textit{provided} that one inserts a weighting factor $(1+\lambda(j_i))^{-1}$ under every sum over a momentum $j_i$, where
\begin{equation}
\lambda(j_1)\equiv\sum_{j_2}(F(j_1,j_2;1,2),\hat{A}^jF(j_1,j_2;1,2))\label{eq.5}
\end{equation}
is assumed small compared to unity. Because we shall evaluate the correlation series exactly only through its $r_s\,\log r_s$ term, where $r_s$ is the applicable Seitz radius, weighting factors may be dropped entirely from the Euler equations. Additionally it is only necessary to include their contributions to the nominally infinite terms of $E_{GSJ}$ (which must sum to zero) and to the HF portion of the kinetic energy, so that Eq.(3) gives\cite{P65, thesis}
\begin{eqnarray}
E_{GSJ}=\sum_{j_1}(1-\lambda(j_1))\left(U(j_1;1),(-\frac{\hbar^2}{2m}\mathbf{\nabla}_1^2)\, U(j_1;1)\right)\nonumber \\
+\frac{1}{2}\sum_{j_1,\,j_2}(1-\lambda(j_1)-\lambda(j_2))\left(U(j_1;1)U(j_2;2),\phi(1,2)U(j_1;1)U(j_2;2)\frac{}{}\right)\frac{}{}\nonumber \\
+\sum_{j_1,\,j_2,\,j_3} \left(\frac{}{}U(j_3;3)F(j_1,j_2;1,2),(\phi(1,3)+\phi(2,3)) U(j_3;3)\hat{A}^jF(j_1,j_2;1,2)\frac{}{}\right)\nonumber \\
+\frac{1}{2}\sum_{j_1,\,j_2}\left(F(j_1,j_2;1,2),-\frac{\hbar^2}{2m}(\mathbf{\nabla}_1^2+\mathbf{\nabla}^2_2)\hat{A}^jF(j_1,j_2;1,2)\frac{}{}\right)\nonumber \\
+\sum_{j_1,\,j_2}\mathsf{Re}\left(U(j_1;1)U(j_2;2),\phi(1,2)\hat{A}^jF(j_1,j_2;1,2)\frac{}{}\right)\nonumber \\  
+\sum_{j_1,\,j_2,\,j_3}\left(\frac{}{}U(j_3;3)F(j_1,j_2;1,2),\phi(2,3)U(j_2;2)F(j_1,j_3;1,3)\right)\nonumber \\
-\sum_{j_1,\,j_2,\,j_3} \left(\frac{}{}U(j_3;3)F(j_1,j_2;1,2),\phi(2,3)(U(j_3;2)F(j_1,j_2;1,3)+U(j_2;3)F(j_1,j_3;1,2))\frac{}{}\right)\nonumber \\
-\frac{1}{2}\sum_{j_1,\,j_2}\left(U(j_1;1)U(j_2;2),\phi(1,2)U(j_2;1)U(j_1;2)\frac{}{}\right)\frac{}{}\nonumber \\ +\frac{1}{2}\sum_{j_1,\,j_2}\left(F(j_1,j_2;1,2),\phi(1,2)F(j_1,j_2;1,2)\right)+E_{background}.\label{eq.6}
\end{eqnarray}
The convention here is that sums run from $1$ to $N$ and that any summand, considered as a function of the indices $j_1,\,j_2,\cdots$, is to be omitted when any two of these indices become equal.

In Eq.(6) the nominally infinite energy associated with the positively charged background $E_b$ reduces the factor weighting the HF Hartree term to just $(-\lambda(j_1)-\lambda(j_2))$ so that this and the first triple sum (also nominally infinite) in Eq.(6) cancel. When one considers infinite terms of next higher order, where the corresponding weighting for the Hartree term is now more accurately $(1-\lambda(j_1)-\lambda(j_2)+\lambda(j_1)\lambda(j_2)-\lambda(j_1)^2-\lambda(j_2)^2)$, with weighting $(1-\lambda(j_1)-\lambda(j_2)-\lambda(j_3))$ applied to the initial triple sum of Eq.(6), and then adding the higher order analogue of this triple summation, (this with a pair of the two electron functions both pre- and post-multiplying a sum of four two body potentials), cancelation again occurs. We shall assume that cancelations continue through all higher orders.

Finally we note that Eq.(6) has been simplified, in comparison to the form which must be used to calculate variations,\cite{P65, thesis} by dropping terms proportionate to $\int d^Dr_1 U^{\dagger}(j_1;1)F(j_1,j_2;1,2)$. It is readily shown that putting this latter integral to zero is consistent with solutions of the following Euler equations.

\subsection*{Euler equations}

Variation with respect to $U^{\dagger}(j_1;1)$ of $E_{GSJ}$ when just the initial two terms of the GSJ series are inserted into Eq.(3) gives the complicated one electron Euler equation discussed in connection with an order $r_s$ term below. Because we do not attempt to evaluate the entirety of contributions to the correlation series order $r_s$ term, it is sufficient to approximate the one electron equation by the applicable form of the HF relation,
\begin{equation}
(\epsilon(j_1)+\frac{\hbar^2}{2m}\mathbf{\nabla}_1^2)\,U(j_1;1)=-\sum_{j_3}\int d^Dr_3 U^{\dagger}(j_3;3)\phi(1.3)U(j_3;1)U(j_1;3).\label{eq.7}
\end{equation}

Variation with respect to $f^{\dagger}(j_1,j_2;1,2)$ of the initial two terms of the GSJ series in Eq.(3) leads to an integro-differential equation whose differential operators act on the combination $(U(j_1;1)U(j_2;2)+f(j_1,j_2;1,2))$ of one and two electron functions. This, together with Eq.(6), can be manipulated to give the two electron Euler equation\cite{P65}
\begin{eqnarray}
(\epsilon(j_1)+\epsilon(j_2)-\frac{\hbar^2}{2m}(\mathbf{\nabla}_1^2 +\mathbf{\nabla}_2^2))F(j_1,j_2;1,2)\nonumber \\
-\sum_{j_3}\int d^D r_3U^{\dagger}(j_3;3)(\tilde{P}(1)\phi(1,3)U(j_1;1)F(j_2,j_3;2,3)+\tilde{P}(2)\phi(2,3)U(j_2;2)F(j_1,j_3;1,3))\nonumber \\
-\tilde{P}(1)\tilde{P}(2)U(j_1;1)U(j_2;2)=\tilde{P}(1)\tilde{P}(2)F(j_1,j_2;1,2)\nonumber \\
-\sum_{j_3}\int d^D r_3U^{\dagger}(j_3;3)\left(\phi(1,3)U(j_3;1)F(j_2,j_1;2,3)+\tilde{P}(1)\phi(1,3)U(j_1;3)F(j_3,j_2;1,2)\right.\nonumber \\
+\left.\phi(2,3)U(j_3;2)F(j_1,j_2;1,3)+\tilde{P}(2)\phi(2,3)U(j_2;3)F(j_1,j_3;1,2)\right),\quad\label{eq.8}
\end{eqnarray}
where the idempotent projection operators  $\tilde{P}(1)\equiv 1-\sum_{j_1^{\prime}}\int d^D r^{\prime}_1U^{\dagger}(j^{\prime}_1;1^{\prime})\tilde{\bigcirc}(1 \!\!\rightarrow\! 1^{\prime})$, with $\tilde{\bigcirc}(1 \!\!\rightarrow\! 1^{\prime})$ operating to replace functions of $\mathbf{r}_1$ by the same function of  $\mathbf{r}^{\prime}_1$, subtract from an arbitrary function that part which is orthogonal to each of the one electron functions. 

Eq.(8) has been arranged with direct terms on the left, while on the right are terms diminished by an additional projection operator or by an exchange of indices $j_2\leftarrow\!\!\!\!\rightarrow j_3$. Still smaller terms, with exchanges involving three indices, are negligible.

\subsection*{Approximating the two electron functions}

For the spatially homogeneous electron gas $\phi(1,2)=e^2/|\mathbf{r}_1-\mathbf{r}_2|$, and the one electron functions are plane waves satisfying periodic boundary conditions over a $D=2$ or $D=3$ dimensional region of side $L$,
\begin{equation}
U(j_1;1)=\frac{1}{\sqrt{L^D}}\exp{\left(\imath K_D\mathbf{k}_1\cdot \mathbf{r}_1\right)}\sigma(j_1).\label{eq.9}
\end{equation}
Here the Fermi wave number $K_D$ is $K_2=\sqrt{2\pi N/L^2}$ in two dimensions and $K_3=\sqrt[3]{3\pi^2N/L^3}$ in three dimensions and $\mathbf{k}_1$ denotes the dimensionless wavenumber $\mathbf{k}(j_1)$. It is convenient to take $\sigma(j)=\uparrow$ for $1\leq j\leq[N/2]$ and $\sigma(j)=\downarrow$ for $[N/2]<j\leq[N]$.

Writing the two electron functions in terms of the real-valued Fourier-transforms ${\cal F}(\mathbf{k}_1,\mathbf{k}_2;\mathbf{q})$,
\begin{equation}
F(j_1,j_2;1,2)=U(j_1;1)U(j_2;2)\int d^D q \exp(\imath K_D\,\mathbf{q}\cdot (\mathbf{r}_1-\mathbf{r}_2)){\cal F}(\mathbf{k}_1,\mathbf{k}_2;\mathbf{q}),\label{eq.10}
\end{equation}
and substituting from Eq.(7) for the Lagrange multipliers, Eq.(8) becomes
\begin{eqnarray}
((\mathbf{k}_1-\mathbf{k}_2)\cdot \hat{\mathbf{e}}_{\mathbf{q}}+q){\cal F}(\mathbf{k}_1,\mathbf{k}_2;\mathbf{q}) \nonumber \\
+\int d^D k_3\Theta(1-k_3)(\Theta(|\,\mathbf{q}+\mathbf{k}_1|-1){\cal F}(\mathbf{k}_3,\mathbf{k}_2;\mathbf{q})
+\Theta(|\,\mathbf{q}-\mathbf{k}_2|-1){\cal F}(\mathbf{k}_1,\mathbf{k}_3;\mathbf{q}) )\nonumber \\
=-\frac{\alpha_D R_D}{\pi^{D-1}q^D}\frac{1}{2}\Theta(|\,\mathbf{q}+\mathbf{k}_1|-1)\Theta(|\,\mathbf{q}-\mathbf{k}_2|-1)\nonumber \\
+\frac{\alpha_D R_D}{\pi^{D-1}q^D}\left[-\frac{1}{2}\Theta(|\,\mathbf{q}+\mathbf{k}_1|-1)\Theta(|\,\mathbf{q}-\mathbf{k}_2|-1)
\int\frac{d^D q^{\prime}}{|\mathbf{q}-\mathbf{q}^{\prime}|^{D-1}} {\cal F}(\mathbf{k_1},\mathbf{k}_2;\mathbf{q}^{\prime})\right.\nonumber \\
+\left.\int d^D k_3\Theta(1-k_3)\left(( \frac{1}{|\mathbf{q}+\mathbf{k}_1+\mathbf{k}_3|^{D-1}}-\frac{1}{|\mathbf{k}_1+\mathbf{k}_3|^{D-1}}) {\cal F}(\mathbf{k}_1,\mathbf{k}_2;\mathbf{q})\right.\right.\nonumber \\
+\left.\left.\frac{\Theta(|\mathbf{q}+\mathbf{k}_1|-1)}{|\mathbf{k}_1-\mathbf{k}_3|^{D-1}}{\cal F}(\mathbf{k}_3,\mathbf{k}_2;\mathbf{q})\right)\right.\nonumber \\
+\left.\int d^D k_3\Theta(1-k_3)\left(( \frac{1}{|\mathbf{q}-\mathbf{k}_2+\mathbf{k}_3|^{D-1}}-\frac{1}{|\mathbf{k}_2-\mathbf{k}_3|^{D-1}}) {\cal F}(\mathbf{k}_1,\mathbf{k}_2;\mathbf{q})\right.\right.\nonumber \\
+\left.\left.\frac{\Theta(|\mathbf{q}-\mathbf{k}_2|-1)}{|\mathbf{k}_2+\mathbf{k}_3|^{D-1}}{\cal F}(\mathbf{k}_1,\mathbf{k}_3;\mathbf{q})\right)\right].\quad \label{eq.11}
\end{eqnarray}
Here $\hat{\mathbf{e}}_{\mathbf{q}}=\mathbf{q}/q$, $\Theta$ is the unit step function, $\alpha_2=1/\sqrt{2}$, $\alpha_3=\sqrt[3]{4/9\pi}$, and the Seitz radius $r_s$ is denoted by $R_2=(\alpha_2 K_2 (\hbar^2/2m))^{-1}$ in two dimensions and $R_3=(\alpha_3 K_3 (\hbar^2/2m))^{-1}$ in three dimensions.

As in Eq.(8) the direct terms on the left in Eq.(11) are large in comparison with the homogeneous terms on the right. Neglecting these latter terms gives a useful first approximation ${\cal F}\approx {\cal F}^{(1)}$ which can be found by setting just the initial terms on left and right in Eq.(11) equal, so that
\begin{equation*}
{\cal F}(\mathbf{k}_1,\mathbf{k}_2;\mathbf{q})\approx\;\approx -\frac{\alpha_D R_{D}}{2(\pi q)^{D-1}} \frac{\Theta(|\,\mathbf{q}+\mathbf{k}_1|-1)\Theta(|\,\mathbf{q}-\mathbf{k}_2|-1)}{(\mathbf{k}_1-\mathbf{k}_2)\cdot \mathbf{q}+q^2}.
\end{equation*} 
This can be written as
\begin{eqnarray}
{\cal F}(\mathbf{k}_1,\mathbf{k}_2;\mathbf{q})\approx\;\approx -\frac{\alpha_D R_{D}}{2\pi^{D-1}q^D}\Theta(|\,\mathbf{q}+\mathbf{k}_1|-1)\Theta(|\,\mathbf{q}-\mathbf{k}_2|-1)\nonumber \\
\times\mathsf{Re}\int_{-\infty}^{\infty}\frac{d\nu}{2\pi\imath}\frac{1}{-\nu+\mathbf{k}_1\cdot\mathbf{e}_{\mathbf{q}}+q/2-\imath \eta}\,\frac{1}{\nu-\mathbf{k}_2\cdot\mathbf{e}_{\mathbf{q}}+q/2-\imath \eta}\,{\cal G}(\nu,q), \label{eq.12}
\end{eqnarray}
where $\eta$ is a positive infinitesimal and where ${\cal G}(\nu,q)=1$. Next substituting Eq.(12), now with ${\cal G}(\nu,q)$ undetermined, into the left hand member of Eq.(11), where the coefficient of ${\cal F}(\mathbf{k}_1,\mathbf{k}_2;\mathbf{q})$ has been written as $(-\nu+\mathbf{k}_1\cdot\mathbf{e}_{\mathbf{q}}+q/2)+(\nu-\mathbf{k}_2\cdot\mathbf{e}_{\mathbf{q}}+q/2)$, and equating the result to zero leads to
\begin{equation}
\mathsf{Re}\int_{-\infty}^{\infty}\frac{d\nu}{2\pi\imath}\left(\frac{1}{-\nu+\mathbf{k}_1\cdot\mathbf{e}_{\mathbf{q}}+q/2-\imath \eta}{\cal D}(-\nu,q)
+\frac{1}{\nu-\mathbf{k}_2\cdot\mathbf{e}_{\mathbf{q}}+q/2-\imath \eta}{\cal D}(\nu,q)\right){\cal G}(\nu,q)=1 \label{eq.13}
\end{equation}
for the determination of ${\cal G}$. Here it is convenient to put
\begin{equation}
{\cal D}(\nu,q)= 1+\frac{R_D}{q^{D-1}}{\cal U}(\nu,q),\label{eq.14}
\end{equation}
with ${\cal U}(\nu,q)$ given by
\begin{equation}
{\cal U}(\nu,q)=\frac{\alpha_D}{\pi^{D-1}q}\int d^D k_3\frac{\Theta(1-k_3)\Theta(|\mathbf{q}+\mathbf{k}_3|-1)}{-\nu+\mathbf{k}_3\cdot\mathbf{e}_{\mathbf{q}}+q/2-\imath \eta}.\label{eq.15}
\end{equation}

The locations of the poles in the integrands of each of the terms on the left of Eq.(13) can be restricted to just one of the half $\nu$-planes by choosing
\begin{equation}
{\cal G}(\nu,q)=({\cal D}(\nu,q){\cal D}(-\nu,q))^{-1},\label{eq.16}
\end{equation}
so that, on account of its symmetry, Eq.(13) will be satisfied provided that
\begin{equation}
\mathsf{Re}\int_{-\infty}^{\infty}\frac{d\nu}{2\pi\imath}\,\frac{1}{-\nu+\mathbf{k}_1\cdot\mathbf{e}_{\mathbf{q}}+q/2-\imath \eta} \,\frac{1}{{\cal D}(\nu,q)} =\frac{1}{2}.\label{eq.17}
\end{equation}
The validity of Eq.(17) rests on assumptions that the reciprocal of ${\cal D}(\nu,q)$ can be expanded in powers of $R_D$ and that, in each term of the resulting series, the integration over $\nu$ can be performed first. Then the first term of the series integrates to one half while the remaining terms vanish since their integrands decrease at least as rapidly as $|\nu|^{-2}$, for large $|\nu|$, and are free of poles in the upper half plane.

Combining Eqs.(12) and (16) gives
\begin{eqnarray}
{\cal F}^{(1)}(\mathbf{k}_1,\mathbf{k}_2;\mathbf{q})=-\frac{\alpha_D R_{D}}{2\pi^{D-1}q^D}\Theta(|\,\mathbf{q}+\mathbf{k}_1|-1)\Theta(|\,\mathbf{q}-\mathbf{k}_2|-1)\nonumber \\ \times\mathsf{Re}\int_{-\infty}^{\infty}\frac{d\nu}{2\pi\imath}\frac{1}{-\nu+\mathbf{k}_1\cdot\mathbf{e}_{\mathbf{q}}+q/2-\imath \eta}\,\frac{1}{\nu-\mathbf{k}_2\cdot\mathbf{e}_{\mathbf{q}}+q/2-\imath \eta}\frac{1}{{\cal D}(\nu,q){\cal D}(-\nu,q)}.\label{eq.18}
\end{eqnarray}

Replacing ${\cal F}$ on the right in Eq.(11) by ${\cal F}^{(1)}$ and on the left by ${\cal F}^{(1)}+{\cal F}^{(2)}$, determines a second order approximation which is discussed below in connection with $R_3\,\log\,R_3$ term of the correlation energy.

\subsection*{Correlation energy}

The correlation energy $\epsilon_D$, again in both two and three dimensions, is obtained from Eq.(6) upon subtracting the HF contributions and dividing by $N$. Similar to the relative importance of terms in Eq.(8), we expect that the direct terms of $\epsilon_D$ will give contributions to the correlation energy of all orders, while terms diminished by exchange, or in the case of the last double sum of Eq.(6), by the implicit presence of an extra pair of projection operators, will give contributions to orders $R_D\,\log\,R_D$ and higher. Identifying these contributions as $\epsilon_D^{(1)}$ and $\epsilon_D^{(2)}$ respectively, and expressing the results in terms of the transformed two electron functions Eq.(6) gives\cite{N3}
\begin{eqnarray}
\epsilon_D^{(1)}({\cal F})=\frac{D}{2^{D-2}\pi (\alpha_D R_D)^2} \int d^Dk_1 \Theta(1-k_1)\int d^D k_2 \Theta(1-k_2)\int d^Dq\,((\mathbf{k}_1+\mathbf{k}_2)\cdot\mathbf{q}+q^2) \nonumber \\
\times\left({\cal F}(\mathbf{k}_1,-\mathbf{k}_2;\mathbf{q})^2-\frac{1}{2}{\cal F}(\mathbf{k}_1,-\mathbf{k}_2;\mathbf{q}){\cal F}(-\mathbf{k}_2,\mathbf{k}_1;\mathbf{q}+\mathbf{k}_1+\mathbf{k}_2)\right)\nonumber \\
+\frac{D}{2^{D-2}\pi^D\alpha_D R_D} \int d^Dk_1 \Theta(1-k_1)\int d^D k_2 \Theta(1-k_2) \nonumber \\
\times\int d^Dq\left(\frac{{\cal F}(\mathbf{k}_1,-\mathbf{k}_2;\mathbf{q})}{q^{D-1}}-\frac{1}{2}\frac{{\cal F}(-\mathbf{k}_2,\mathbf{k}_1;\mathbf{q}+\mathbf{k}_1
+\mathbf{k}_2)}{|\mathbf{q}+\mathbf{k}_1+\mathbf{k}_2|^{D-1}}\right.\nonumber \\
+\left.\frac{2}{q^{D-1}}\int d^D k_3 \Theta(1-k_3){\cal F}(\mathbf{k}_1,-\mathbf{k}_2;\mathbf{q}){\cal F}(\mathbf{k}_1,\mathbf{k}_3;\mathbf{q})\right),\qquad \label{eq.19}
\end{eqnarray}
and
\begin{eqnarray}
\epsilon_D^{(2)}({\cal F})=\frac{D}{2^{D-2}\pi^D \alpha_D R_D}\int d^Dk_1 \Theta(1-k_1)\int d^D k_2 \Theta(1-k_2)\int d^Dq_1 {\cal F}(\mathbf{k}_1,-\mathbf{k}_2;\mathbf{q}_1)\nonumber \\
\times\left[-\int d^D k_3 \Theta(1-k_3)\left((\frac{1}{|\mathbf{q}_1+\mathbf{k}_1+\mathbf{k}_3|^{D-1}}-\frac{1}{|\mathbf{k}_1+\mathbf{k}_3|^{D-1}}){\cal F}(\mathbf{k}_1,-\mathbf{k}_2;\mathbf{q}_1)\right.\right.\nonumber \\
\left.\left.+\frac{{\cal F}(\mathbf{k}_2,-\mathbf{k}_3;\mathbf{q}_1)}{|\mathbf{k}_1-\mathbf{k}_3|^{D-1}}\right)+\frac{1}{2}\int d^Dq_2 \frac{{\cal F}(\mathbf{k}_1,-\mathbf{k}_2;\mathbf{q}_2)}{|\mathbf{q}_1-\mathbf{q}_2|^{D-1}}\right],\quad\label{eq.20}
\end{eqnarray}
where the units are Ry rather than ergs. Note that the $\lambda(j_1)$ dependent reduction of the HF kinetic energy term in Eq.(6) contributes to the kinetic energy in Eq.(19).

Substituting Eq.(18) in (19), and using the assumptions which underlie Eq.(17) to evaluate one of the two contour integrations in the two direct bilinear terms of Eq.(19), gives
\begin{eqnarray}
\epsilon_D^{(1)}({\cal F}^{(1)})=\frac{D}{2\pi\alpha_D^2}\int_0^{\infty}d\nu\int_0^{\infty}\frac{dq}{q^{D-2}}\mathsf{Im}\left(\frac{{\cal U}(\nu,q){\cal U}(-\nu,q)}{|{\cal D}(\nu,q)|^2{\cal D}(-\nu,q)^2}-2\frac{{\cal U}(\nu,q){\cal U}(-\nu,q)}{{\cal D}(\nu,q){\cal D}(-\nu,q)}\right.\nonumber \\
\left.+\frac{R_D}{q^{D-1}}\frac{{\cal U}(\nu,q){\cal U}(-\nu,q)^2}{|{\cal D}(\nu,q)|^2{\cal D}(-\nu,q)^2}\right)
+\left\{
\begin{array}{c}
0.2039\\
0.0484
\end{array} 
\right\}\label{eq.21}
\end{eqnarray}
which is equivalent to
\begin{equation}
\epsilon_D^{(1)}({\cal F}^{(1)})=-\frac{D}{2\pi\alpha^2}\int_0^{\infty}d\nu\int_0^{\infty}\frac{dq}{q^{D-2}}\mathsf{Im}\left(\frac{{\cal U}(\nu,q){\cal U}(-\nu,q)}{{\cal D}(\nu,q){\cal D}(-\nu,q)}
\right)
+\left\{
\begin{array}{c}
0.2039\\
0.0484
\end{array} 
\right\},
\end{equation} \label{eq.22}
where we have used the fact that the imaginary part of the function ${\cal U}(\nu,q)$, evaluated in Appendix A,
vanishes for $\nu<0$.

The additive constants (second order Rayleigh-Schrodinger exchange energies) in Eqs.(21) and (22) are the sum of the kinetic and potential energy exchange terms of Eq.(19) in the limit $R_D\rightarrow 0$. These were evaluated  in cylindrical coordinates in two dimensions (or Cartesian coordinates in three dimensions) with symmetry axis $z$ along $\mathbf{q}$ and with the common origin of the wavenumber vectors at  $z=q/2$. A 45 degree rotation of axes allows reductions to three-fold quadratures in both cases with results in agreement with those of previous authors.\cite{Onsager, RK}

In two dimensions, development of Eq.(22) in partial fractions now yields a constant, $-0.4039$ Ry, plus terms proportional to $R_2\,\log R_2$ and to $R_2$. Numerical fitting at 100 points $R_2^{(k)}=10^{0.05k-8};\,k=1,\cdots 100$, so that $R_2<10^{-3}$ where the contribution from order $R_2^2$ terms is expected to be small, gives
\begin{equation}
\epsilon_2^{(1)}({\cal F}^{(1)})=-0.4039-0.50\,R_2\,\log R_2-0.87\,R_2, \label{eq.23}
\end{equation}
with an RMS deviation of $1.1\,10^{-5}$ from computed values. The result here is in agreement with that found by Rajagopal and Kimball\cite{RK} with the exception of the $R_2\,\log\,R_2$ term which is approximately three times larger. 

In three dimensions the presence of a logarithmic divergence with small $R_3$ makes direct fitting unfeasible.  Instead means must be found to subtract the logarithmic dependence prior to evaluation. One way to do this is to integrate in Eq.(21) with respect to only its \textit{explicit} dependence on the momentum $q$. This renders the integrand in Eq.(22) in the form of a partial derivative with respect to $q$, of a quantity proportionate to
\begin{equation}
G(\nu,q)\equiv\frac{{\cal U}(\nu,q){\cal U}(-\nu,q)}{{\cal U}(-\nu,q)-{\cal U}(\nu,q)}({\cal U}(\nu,q)\log(q^2 +R_3\,{\cal U}(\nu,q))-{\cal U}(-\nu,q)\log(q^2 +R_3\,{\cal U}(-\nu,q))).\label{eq.24}
\end{equation} 
Expressing this partial derivative in terms of its related total derivative isolates the divergence at small $R_3$.

A complication of this method is that the imaginary part of the integrand in Eq.(22) picks up contributions from a new area in the first quadrant of the $\nu-q$ plane where the argument of $\log(q^2 +R_3\,{\cal U}(\nu,q))$ in Eq.(24) becomes negative. This happens on and to the right of the line $q=2(\nu-1)$ and below the curve $q=q_c(\nu)$, where $q_c(\nu)^2+R_3\,{\cal U}(\nu,q_c(\nu))=0$.  Here $q_c(\nu)$ originates at a point $0<\nu<1$ on the $\nu$ axis, reaches a maximum for $\nu \stackrel{_>}{_\approx} 1$, and, with increasing $\nu$, turns downward, crossing the line $q=2(\nu-1)$ at the point $(\nu_c,2(\nu_c-1))$, where $\nu_c\geq 1$ for $R_3\geq 0$. Finally $q_c(\nu)$ approaches zero for large $\nu$.

Summing contributions in Eq.(22) from the area between the lines $q=2(\nu\pm 1)$, where we have $\mathsf{Im}\,{\cal U}(\nu,q)\neq 0$, and then from the area where $\mathsf{Im}\,\log(q^2 +R_3\,{\cal U}(\nu,q))=\pi$, gives
\begin{eqnarray}
\epsilon_3^{(1)}=\frac{3}{\pi\alpha^2}\left[\left(\int_0^1d\nu\int_0^{2(\nu+1)}dq+\int_1^{\infty}d\nu\int_{2(\nu-1)}^{2(\nu+1)}dq\right)\right.\ \nonumber \\
\left.\times\left(\frac{d}{dq}-\frac{\partial\,{\cal U}(\nu,q)}{\partial q}\frac{\partial}{\partial\,{\cal U}(\nu,q)}-\frac{\partial\,{\cal U}(-\nu,q)}{\partial q}\frac{\partial}{\partial\,{\cal U}(-\nu,q)}\right)\mathsf{Im}\,G(\nu,q)\right.\nonumber \\
+\left.\int_1^{\infty}d\nu\int_0^{2(\nu-1)}dq\frac{{\cal U}(\nu,q)^2{\cal U}(-\nu,q)}{{\cal U}(-\nu,q)-{\cal U}(\nu,q)}\frac{\partial}{\partial q}\pi \Theta(-(q^2+R_3\,{\cal U}(\nu,q)))\right].\quad\label{eq.25}
\end{eqnarray}
Here upper limits of the $q$ integrations over the total derivative do not contribute, while the lower limit of the first of these integrations is proportionate to $\log\,R_3$. The lower limit of the second $q$ integration over the total derivative leaves an integration over $\nu$ running from one to infinity with $\mathsf{Im}\,G(\nu,q)$ evaluated on the right most bounding line $q=2(\nu-1)$. However only along the initial portion of this bounding line, where $\nu<\nu_c$ so that $q^2+R_3\,{\cal U}(\nu,q)<0$, is $\mathsf{Im}\,G(\nu,q)$ non-zero. The sum of the double integrals over the partial derivatives of $\mathsf{Im}\,G(\nu,q)$ is bounded.

In the final double integral in Eq.(25) the derivative of the step function is proportional to $\delta(q-q_c(\nu))$ so the integral over $q$ vanishes where $2(\nu-1)<q_c(\nu)$, that is where $\nu<\nu_c$. This leaves for the last line of Eq.(25)
\begin{equation}
-\frac{3}{4\alpha_3^2}\int^{\infty}_{\nu_c}d\nu\frac{{\cal U}(\nu,q_c(\nu))^2{\cal U}(-\nu,q_c(\nu))}{{\cal U}(-\nu,q_c(\nu))-{\cal U}(\nu,q_c(\nu))}\,\frac{1}{1+(R_3/2q_c(\nu))(\partial/\partial q_c(\nu)){\cal U}(\nu,q_c(\nu))},\nonumber
\end{equation}
where the final fraction approaches unity as $R_3$ approaches zero since the $q$ derivative of $\mathsf{Re}\;{\cal U}(\nu,q)$ is itself proportional to $q$.

With the divergent term, $0.0622\,\log\,R_3$ Ry, set aside, the remainder was examined numerically at 60 points $R_3^{(k)}=10^{0.05k-6};\,k=1,\cdots 60$, so that $0< R_3<0.001$, with the result, inclusive of the second order exchange energy $0.0484$ Ry,
\begin{equation}
\epsilon_3^{(1)}({\cal F}^{(1)})=0.0622\,\log R_3-0.082-0.015\,R_3, \label{eq.26}
\end{equation}
with an RMS deviation of $1.2\,10^{-5}$ from computed values. No $\,R_3\log R_3$ dependence was found. We note that if the direct potential energy term quadratic in the two electron functions in Eq.(8), (this is responsible for the term proportionate to $R_3$ on the right in Eq.(21)), were not included in the subsequent calculations, the constant term of Eq.(26) becomes $-0.093$ Ry, approximately the result found from perturbation theory.\cite{GB}

\subsection*{Order $R_D\,\log\,R_D$}

Logarithmic contributions are determined in the transformed domain by behavior at small $q$, where the functions ${\cal U}$ in the denominator of Eq.(18) approach finite positive limits. Accordingly we may now simplify ${\cal F}^{(1)}$ by replacing each of the functions ${\cal D}$ in the denominator of Eq.(18) by $(1+
R_D/q^{D-1})$ and performing the $\nu$-integration to get
\begin{equation}
{\cal F}^{(1)}(\mathbf{k}_1,\mathbf{k}_2;\mathbf{q})\approx -\frac{\alpha_D R_{D}}{2\pi^{D-1}q^D}\frac{\Theta(|\,\mathbf{q}+\mathbf{k}_1|-1)\Theta(|\,\mathbf{q}-\mathbf{k}_2|-1)} {(1+
R_D/q^{D-1})^2((\mathbf{k}_1-\mathbf{k}_2)\cdot \mathbf{\hat{e}}_{\mathbf{q}}+q)}.\label{eq.27}
\end{equation}
In terms of this simplified version of ${\cal F}^{(1)}$, we have
\begin{eqnarray}
\epsilon_D^{(2)}({\cal F}^{(1)})=\frac{D\alpha_D R_D}{2^D\pi^{3D-2}}\left\{-\int \frac{d^D q}{q^{2D}}\frac{1}{(1+R_D/q^{D-1})^4}\int d^D k_1\Theta(1-k_1)\Theta(|\mathbf{q}+\mathbf{k}_1|-1)\right.\nonumber \\
\left.\times\int d^D k_2 \Theta(1-k_2)  \frac{\Theta(|\mathbf{q}+\mathbf{k}_2|-1)}{((\mathbf{k}_1+\mathbf{k}_2)\cdot\hat{\mathbf{e}}_{\mathbf{q}}+q)}\left[\frac{1}{(\mathbf{k}_1+\mathbf{k}_2)\cdot\hat{\mathbf{e}}_{\mathbf{q}}+q}\int d^D k_3\,\Theta(1-k_3) \right.\right.\nonumber \\
\times\left.\left. \left(\frac{1}{|\mathbf{q}+\mathbf{k}_1+\mathbf{k}_3|^{D-1}}-\frac{1}{|\mathbf{k}_1+\mathbf{k}_3|^{D-1}}\right)+\int d^D k_3\frac{\Theta(1-k_3)\, \Theta(|\mathbf{q}+\mathbf{k}_3|-1)}{((\mathbf{k}_2+\mathbf{k}_3)\cdot\hat{\mathbf{e}}_{\mathbf{q}}+q)\;|\mathbf{k}_1-\mathbf{k}_3|^{D-1}}\right]\right.\nonumber \\
+\left.\frac{1}{2}\int \frac{d^D q_1}{q^D_1}\frac{1}{(1+R_D/q_1^{D-1})^2}\int\frac{d^D q_2}{q^D_2}\frac{1}{(1+R_D/q^{D-1}_2)^2} 
\frac{1}{|\mathbf{q}_1-\mathbf{q}_2|^{D-1}}\int d^D k_1 d^D k_2\Theta(1-k_1)\right.\nonumber \\
\left.\times \Theta(1-k_2) \frac{\Theta(|\mathbf{q}_1+\mathbf{k}_1|-1)\Theta(|\mathbf{q}_1+\mathbf{k}_2|-1)}{(\mathbf{k}_1+\mathbf{k}_2)\cdot\mathbf{e}_{\mathbf{q}_1}+q_1}\, \frac{\Theta(|\mathbf{q}_2+\mathbf{k}_1|-1) \Theta(|\mathbf{q}_2+\mathbf{k}_2|-1)}{(\mathbf{k}_1+\mathbf{k}_2)\cdot\mathbf{e}_{\mathbf{q}_2}+q_2}\right\}.\qquad\label{eq.28}
\end{eqnarray}

In two dimensions the curly-bracketed term of Eq.(28) remains finite in the limit $R_2\rightarrow 0$ so that there is no order $R_2\,\log R_2$ contribution. In three dimensions the $q$-integrations over the square bracketed integrals in Eq.(28), considered in isolation, each diverge logarithmically. However, as shown in Appendix B, these divergences cancel (as they must) with the result that
\begin{equation}
\epsilon_3^{(2)}=0.059\,R_3\,\log\,R_3. \label{eq.29}
\end{equation}

A second potential source of order $R_D\,\log\,R_D$ contributions is
\begin{eqnarray}
\epsilon_D^{(1)}({\cal F}^{(2)}+{\cal F}^{(1)})-\epsilon_D^{(1)}({\cal F}^{(1)})=\frac{D}{(2\pi)^{D-2}(\alpha_D R_D)^2}
\int d^Dk_1 \Theta(1-k_1)\int d^D k_2 \Theta(1-k_2) \nonumber \\
\times\int d^Dq((\mathbf{k}_1+\mathbf{k}_2)\cdot\mathbf{q}+q^2){\cal F}^{(1)}(\mathbf{k}_1,-\mathbf{k}_2;\mathbf{q}){\cal F}^{(2)}(\mathbf{k}_1,-\mathbf{k}_2;\mathbf{q})\nonumber \\
+\frac{D}{\pi^D\alpha_D R_D}\int d^Dk_1 \Theta(1-k_1)\int d^D k_2 \Theta(1-k_2)\int \frac{d^Dq}{q^{D-1}} {\cal F}^{(2)}(\mathbf{k}_1,-\mathbf{k}_2;\mathbf{q}),\qquad\label{eq.30}
\end{eqnarray}
where the small exchange and final terms of Eq.(19) have been dropped. Here we expect
${\cal F}^{(2)}$ to be well defined as a result of cancelations of the small $q$ divergences on the right in Eq.(11) analogous to the cancelation observed in Eq.(28).

Expressing the left hand member of Eq.(11) in the form of an integral operator $\hat{B}(\mathbf{k}_1,\mathbf{k}_2;\mathbf{q})$ acting on ${\cal F}$, then, analogous with the classical case,\cite{MT} the inverse operator
\begin{equation}
\hat{B}(\mathbf{k}_1,\mathbf{k}_2;\mathbf{q})^{-1}=\mathsf{Re}\int_{-\infty}^{\infty}\frac{d\nu}{2\pi\imath}\hat{V}(\mathbf{k}_1,\mathbf{q},\nu) \hat{V}(\mathbf{k}_2,-\mathbf{q},-\nu),\label{eq.31}
\end{equation}
where
\begin{eqnarray}
\hat{V}(\mathbf{k}_1,\mathbf{q},\nu)=\frac{1}{-\nu+\mathbf{k}\cdot\mathbf{\hat{e}}_{\mathbf{q}}+q/2-\imath \eta}\nonumber \\
\times\left(1-\frac{\alpha_D R_D}{\pi^{D-1}q^D}\frac{\Theta(|\mathbf{q}+\mathbf{k}|-1)}{{\cal D}(\nu,q)}\int d\mathbf{k}^{\prime}\frac{\Theta(1-k^{\prime})}{-\nu+\mathbf{k}^{\prime}\cdot\mathbf{\hat{e}}_{\mathbf{q}}+q/2-\imath \eta}\hat{\bigcirc}(\mathbf{k}\rightarrow\mathbf{k}^{\prime})\right),\nonumber
\end{eqnarray}
allows one to obtain an explicit expression for ${\cal F}^{(2)}(\mathbf{k}_1,\mathbf{k}_2;\mathbf{q})$.
This, however, turns out not to be needed since substitution from Eq.(27) shows that the order $R_D\,\log\,R_D$ terms on the right in Eq.(30) sum to zero in both two and three dimensions. 

Summing Eqs.(26) and (29) gives finally
\begin{equation}
\epsilon_3=0.0622\,\log\,R_3-0.082+0.059\,R_3\,\log R_3-0.015\,R_3.\label{eq.32}
\end{equation}

\subsection*{Order $R_3$}

An accounting of contributions to order $R_3$ of $\epsilon_3$ requires use of the initial \textit{two} terms of the GSJ series in Eq.(3) when deriving the one electron Euler equation with the consequent inclusion on the right in Eq.(7) of the additional terms\cite{thesis}
\begin{eqnarray}
+\sum_{j_3}\int d^3r_3 U^{\dagger}(j_3;3)\phi(1.3)F(j_1,j_2;1,2)\nonumber \\
-\sum_{j_3,\,j_4}\int d^3r_3d^3r_4U^{\dagger}(j_3;3)U^{\dagger}(j_4;4)\phi(3,4)U(j_1;3)F(j_4,j_3;1,4).\label{eq.33}
\end{eqnarray} 
(It is still unnecessary to include weighting functions.) Like the HF exchange term, these latter two terms, when imported into the two electron equation via the Lagrange multipliers, diverge with small $q$. As happened with the HF exchange contributions, the divergent portions of these contributions are canceled by four additional bilinear terms on the right in Eq.(8),
\begin{eqnarray}
\sum_{j_3,\,j_4}\int d^3r_3d^3r_4F^{\dagger}(j_3,j_4;3,4)\phi(3,4)U(j_4;4)\left(U(j_3;1)F(j_2,j_1;2,3)+U(j_3;2)F(j_1,j_2;1,3)\right)\nonumber \\
-\sum_{j_3,\,j_4,\,j_5}\int d^3r_3d^3r_4d^3r_5 F^{\dagger}(j_3,j_4;3,4)U^{\dagger}(j_5;5)\phi(4,5)U(j_3;5)U(j_4;4)\nonumber \\
\times \left(U(j_5;1)F(j_2,j_1;2,3)+U(j_5;2)F(j_1,j_2;1,3)\right),\qquad\label{eq.34}
\end{eqnarray}
which arise when one extracts, without approximation, a relation for $F(j_1,j_2;1,2)$ alone from the original result obtained from variations, with respect to $f^{\dagger}(j_1,j_2;1,2)$, as discussed following Eq.(7).

Numerical evaluation of all order $R_3$ contributions is problematic since the simplification in Eq.(27) cannot be used. In the transformed domain, order $R_3$ integrands contain two or three additional powers of $q$ in comparison to those of order $R_3\,\log\, R_3$ terms so that short range effects dominate. The alternating signs of these integrals, and the fact that each contains at least one exchange term, leaves $-0.015$ Ry, Eq.(26), from the direct terms of Eq.(8), as a reasonable guess for the order of magnitude of the order $R_3$ contribution to $\epsilon_3$. Thus the correlation energy's order $R_3$ term is found to be roughly equal, and the order $R_3\,\log\,R_3$ term some three times larger, than the respective results obtained from perturbation theory.\cite{Carr}

\setcounter{equation}{0}
\renewcommand{\theequation}{\mbox{A}\arabic{equation}}
\appendix*
\section*{Appendix A}

In the following inequalities are arranged with the expectation that the $\nu$ integrations required by Eq.(25) be performed first.

In two dimensions it is convenient to define the auxiliary functions
\begin{eqnarray}
\begin{array}{l}
L(z,\nu,q)=\sqrt{1-(\nu -q/2)^2}\,\log \left(\frac{f(z,\nu,q)-\sqrt{(1-(\nu-q/2)^2)(1-(z-q/2)^2)}}{z-\nu}\right)+g(z,\nu,q),\nonumber\\
A(z,\nu,q)=\sqrt{(\nu-q/2)^2-1}\,\arcsin\left(\frac{f(z,\nu,q)}{z-\nu}\right)+g(z,\nu,q),\nonumber\\
C(z,\nu,q)=\sqrt{(\nu-q/2)^2-1}\,\arccos\left(\frac{f(z,\nu,q)}{z-\nu}\right)+g(z,\nu,q),\nonumber 
\end{array}
\end{eqnarray}
where
\begin{eqnarray}
\begin{array}{l}
f(z,\nu,q)=1-q^2/4+\nu q/2-\nu z+qz/2,\nonumber\\
g(z,\nu,q)=(\nu -q/2)\arcsin(q/2-z)+\sqrt{1-(z-q/2)^2}.\nonumber
\end{array}
\end{eqnarray}
Next we define
\begin{eqnarray}
\begin{array}{l}
Llt2(\nu,q)=L(1+q/2,\nu,q)-L(0,\nu,q),\nonumber\\
Lgt2(\nu,q)=L(q/2+1,\nu,q)-L(q/2-1,\nu,q),\nonumber\\
Alt2(\nu,q)=A(1+q/2,\nu,q)-A(0,\nu,q),\nonumber\\
Agt2(\nu,q)=A(q/2+1,\nu,q)-A(q/2-1,\nu,q),\nonumber\\
Clt2(\nu,q)=C(1+q/2,\nu,q)-C(0,\nu,q),\nonumber\\
Cgt2(\nu,q)=C(q/2+1,\nu,q)-C(q/2-1,\nu,q).\nonumber
\end{array}
\end{eqnarray}

In terms of this last set of definitions
\begin{eqnarray}
  \mathsf{Re}\,{\cal U}(q,\nu)=
   \frac{2\alpha_2}{\pi q}
     \mathsf{Re}\left\{ \begin{array} {l}
     Llt2(\nu,q)-Llt2(\nu,-q);\quad -1+q/2<\nu<1+q/2 \text{ and }0<q<2,\nonumber\\
     Slt2(\nu,q)-Slt2(\nu,-q);\quad 1+q/2<\nu \text{ and }0<q<2,\nonumber\\
     Clt2(\nu,q)-Clt2(\nu,-q);\quad \nu<-1+q/2 \text{ and }0<q<2,\nonumber \\
     Lgt2(\nu,q);\quad -1+q/2<\nu<1+q/2 \text{ and }q>2,\nonumber\\
     Sgt2(\nu,q);\quad 1+q/2<\nu \text{ and }q>2,\nonumber\\
     Cgt2(\nu,q);\quad \nu<-1+q/2 \text{ and }q>2,\nonumber \\
     \end{array}  \right.\\ \label{eq.A4}
   \end{eqnarray}
and
\begin{eqnarray}
  \mathsf{Im}\,{\cal U}(q,\nu)=
   {\everymath{\displaystyle}2\alpha_2\left\{ \begin{array} {l}
   \vspace{.1in}
   \frac{2\nu}{\sqrt{1-(\nu-q/2)^2}+\sqrt{1-(\nu+q/2)^2}};\quad 0<\nu<1-q/2\text{ and }0<q<2,\nonumber\\
   \frac{\sqrt{1-(\nu-q/2)^2}}{q};\quad (1-q/2<\nu<1+q/2\text{ and }0<q<2)\nonumber\\
   \qquad\text{ or }(q/2-1<\nu<q/2+1\text{ and }q<2),\\
   0;\quad\text{otherwise}.\label{eq.A5}
     \end{array}  \right.}\\ \label{eq.A2}
\end{eqnarray}

In three dimensions
\begin{eqnarray}
    \mathsf{Re}\,{\cal U}(q,\nu)=
   \frac{\alpha_3}{\pi}\mathsf{Re}{\everymath{\displaystyle}\left\{ \begin{array} {l}
  1-\nu-2\nu\log(\nu)+\frac{1-(\nu-q/2)^2}{q}\log(1-\nu+q/2) \nonumber \\
  \qquad -\frac{1-(\nu+q/2)^2}{q} \log(1-\nu-q/2); \quad 0<q<2,  \nonumber \\
   1-\frac{2\nu}{q}+\frac{1-(\nu-q/2)^2}{q} \log \left(\frac{1-\nu+q/2}{1+\nu-q/2}\right);\quad q>2, 
  \end{array}   \right.}\\ \label{eq.A3}
\end{eqnarray}
and
\begin{eqnarray}
  \mathsf{Im}\,{\cal U}(q,\nu)=
  {\everymath{\displaystyle}\alpha_3 \left\{ \begin{array} {lcl}
  2\nu;\quad 0<\nu<1-q/2 \text{ and } 0<q<2, \nonumber \\
  \frac{(1-(\nu-q/2)^2)}{q}; \quad (1-q/2<\nu<1+q/2\text{ and }0<q<2  ) \nonumber \\
  \qquad \text{ or }(q/2-1<\nu<q/2+1\text{ and }q>2  ), \\
   0;\quad\text{otherwise}.
  \end{array}   \right.}\\ \label{eq.A7}
\end{eqnarray}

\setcounter{equation}{0}
\renewcommand{\theequation}{\mbox{B}\arabic{equation}}
\appendix*
\section*{Appendix B}
We adopt, for the $\mathbf{k}_1$ and $\mathbf{k}_2$ integrations in the first term within the square brackets of Eq.(28), spherical coordinates with $z$ axis parallel to $\mathbf{q}$. For small $q$ the step functions require that $1>k_1>1-q\cos\,\theta_1$, where $\cos\,\theta_1>0$, with a corresponding relation for $k_2$. Carrying out the trivial integrations over $\mathbf{k}_3$, $\phi_1$ and $\phi_2$ gives
\begin{eqnarray}
-\frac{3\alpha R_3}{8\pi^7}(2\pi)^3\int_0 \frac{d\mathbf{q}}{q^6} \int_0^{\pi/2}\sin\theta_1 d\theta_1\int_{1-q\cos\,\theta_1}^1 dk_1\,k_1^2 \int_0^{\pi/2}\sin\theta_2 d\theta_2\int_{1-q\cos\,\theta_2}^1 dk_2\,k_2^2\nonumber \\
\times\frac{1}{(1+R_3/q^2)^4}\frac{1}{(k_1\cos\theta_1+k_2\cos\theta_2+q)^2}\nonumber \\
\times\left(\frac{1-|\mathbf{k}_1+\mathbf{q}|^2}{2|\mathbf{k}_1+\mathbf{q}|}\log\left|\frac{1+|\mathbf{k}_1+\mathbf{q}|}{1-|\mathbf{k}_1+\mathbf{q}|}\right|-\frac{1-k_1^2}{2k_1}\log\left|\frac{1+k_1}{1-k_1}\right|\right). \label{B1}
\end{eqnarray}
Putting $|\mathbf{k}_1+\mathbf{q}|\approx k_1+q\cos\theta_1$ and introducing dummy variables of integration $y_i$ such that $k_i=1-qy_i\cos\theta_i$ for $i=1,2$ gives, again for small $q$,
\begin{eqnarray}
-\frac{12\alpha R_3}{\pi^3}\int_0 dq\frac{q^6}{(q^2+R_3)^4}\int_0^1 dz_1\int_0^1 dy_1\,z_1\int_0^1 dz_2\int_0^1 dy_2\,z_2\frac{1}{(z_1+z_2)^2}\nonumber \\
\times\left(\frac{1-(1-qy_1z_1+qz_1)^2}{2}\log \left|\frac{2}{1-(1-qy_1z_1+qz_1)}\right|\right.\nonumber \\
-\left.\frac{1-(1-qy_1z_1)^2}{2}\log\left|\frac{2}{1-(1-qy_1z_1)}\right|\right), \label{B2}
\end{eqnarray}
where we have denoted the cosines $\cos\theta_i$ by $z_i$ for $i=1,2$.

For the second term within the square brackets of Eq.(28) we adopt cylindrical coordinates, with symmetry axis along $\mathbf{q}$, for the integrations over $\mathbf{k}_i$, $i=1,2,3$. Here the origins of the cylindrical systems are located at a point $-\mathbf{q}/2$ from the origins of the $\mathbf{k}_i$ on the $z$-axis. Thus, if $\mathbf{k}_i$ locates a point $(\rho_i,\,z_i,\,\phi_i)$ then $\mathbf{k}_i\cdot \hat{\mathbf{e}}_{\mathbf{q}}=z_i-q/2$. The small $q$ approximations of the integral operators $\int d\mathbf{k}_i\Theta(1-k_i)\,\Theta(|\mathbf{q}+\mathbf{k}_i|-1)$ now appear as
\begin{equation}
\int_0^1 dz_i\int_{\sqrt{1-(z_i+q/2)^2}}^{\sqrt{1-(z_i-q/2)^2}} \rho_id\rho_i\int_0^{2\pi}d\phi_i;\;i=1,\,2,\,3. \nonumber
\end{equation}

Because $\int_0^{2\pi}d\phi_3(\mathbf{k}_1-\mathbf{k}_3)^{-2}$ is a function of $\rho_1^2$ and $\rho_3^2$ only, it is convenient to define $Y_i=\rho_i^2$ for $i=1,\,2,\,3$. Thus for small $q$ this second term becomes, upon performing the $\phi$-integrations
\begin{eqnarray}
-\frac{3\alpha R_3}{\pi^3}\int_0 dq\frac{q^4}{(q^2+R_3)^4}\int_0^1dz_1\int_0^1dz_2\int_0^1dz_3 \nonumber \\
\times\int_{1-(z_1+q/2)^2}^{1-(z_1-q/2)^2}dY_1\int_{1-(z_2+q/2)^2}^{1-(z_2-q/2)^2}dY_2\int_{1-(z_3+q/2)^2}^{1-(z_3-q/2)^2}dY_3 \nonumber \\
\times\frac{1}{\sqrt{\left(\frac{}{}Y_1+(z_1-q/2)^2+Y_3+(z_3-q/2)^2-2(z_1-q/2)(z_3-q/2)\right)^2-4Y_1Y_3}}. \label{B3}
\end{eqnarray}

A change of dummy variables from the $Y_i$ to $y_i$, where $Y_i=1-z_i^2+z_iy_iq-q^2/4$ for $i=1,\,2,\,3$, allows the interchange of the integrals over $z_3$ and $y_3$, while reducing the argument of the square root in (B3) to a quadratic in $z_3$, with the reciprocal of the square root being multiplied by $z_1z_2z_3/((z_1+z_2)(z_2+z_3))$, so that the $z_3$ integral in (B3) can be evaluated analytically. These changes of variable give a multiplicative factor of $q^3$ so that the lower limit of the initial integration over $q$ yields (for at least the $q$-independent part of its integrand) the expected the $R_3\,\log R_3$ dependence.

The $z_3$-integration reveals a divergent term proportional to $\log\,q$ which cancels the $\log\,q$ divergence found in (B2). The remaining terms form a power series in $q$ whose constant term can be evaluated numerically with the result $0.0172\,R_3\,\log\,R_3$ Ry.

The wave number integrations found in the last term of Eq.(28) are best evaluated in spherical coordinates. Here we observe that, for small $q_1$ and $q_2$, the arguments of the four step functions present can be replaced by $k_i-(1-q_j\cos \theta_{i\,j})$, with $i=1,\,2$ and $j=1,\,2$ as appropriate. Here, for a non-zero result, $\cos \theta_{i\,j}=\mathbf{k}_i\cdot \mathbf{q}_j/(k_i\,q_j)$ must be positive. We next introduce spherical coordinates with $z$-axis along $\mathbf{k}_1$ and $x$-axis in the $\mathbf{k}_1-\mathbf{k}_2$ plane such that the $x$-component of $\mathbf{k}_2$ is positive. The angle $\theta_{21}$ between $\mathbf{k}_2$ and $\mathbf{q}_1$ can be expressed in terms of the polar and azimuthal angles of $\mathbf{q}_1$ by means of the relation $\cos\theta_{21}=\cos\beta\,\cos\theta_{11}+\sin\beta\,\sin\theta_{11}\,\cos\phi_1$, where $\beta$ is the angle separating $\mathbf{k}_1$ and $\mathbf{k}_2$. A corresponding expression gives $\cos\theta_{22}$ in terms of $\beta$, $\theta_{12}$ and $\phi_2$. 

The final term of Eq.(28) now becomes
\begin{eqnarray}
\frac{3\alpha_3 R_3}{4\pi^5}\int^{\pi}_0d\beta\,\sin\beta\int_0 d\mathbf{q_2}\frac{q_2}{(q^2_2+R_3)^2}\int_0 d\mathbf{q}_1 \frac{q_1}{(q^2_1+
r-3)^2}\frac{1}{q_1^2-2\mathbf{q}_1\cdot\mathbf{q}_2+q_2^2}  \nonumber \\
\times \int^1_0 dk_1\,k_1^2\int^1_0 dk_2\,k_2^2\, \frac{\Theta(k_1-(1-q_1\cos \theta_{11}))\,\Theta(\cos\theta_{11})\,\Theta(k_2-(1-q_1\cos \theta_{21}))\,\Theta(\cos\theta_{21})}{k_1\cos\,\theta_{11}+k_2\cos\,\theta_{21}} \nonumber \\
\times \frac{\Theta(k_1-(1-q_2\cos \theta_{12}))\,\Theta(\cos\theta_{12})\,\Theta(k_2-(1-q_2\cos \theta_{22}))\,\Theta(\cos\theta_{22})}{k_1\cos\,\theta_{12}+k_2\cos\,\theta_{22}}, \qquad \label{B4}
\end{eqnarray} 
where the zero subscript marking the $\mathbf{q}$ integrations indicates that only the $q\rightarrow 0$ lower limit is expected to contribute.

The step functions in (B4) require $k_1$ and $k_2$ to differ from unity only by terms of order $q$ so that, with the exception of their appearance in the arguments of the step functions, $k_1$ and $k_2$ in (B4) may each be replaced by unity. The resulting trivial integrals over $k_1$ and $k_2$ give
\begin{eqnarray}
\frac{3\alpha R_3}{4\pi^5}\int^{\pi}_0 d\beta\,\sin\beta\int_0 d \mathbf{q}_2\frac{q_2}{(q_2^2+R_3)^2}\int_0 d \mathbf{q}_1 \frac{q_1}{(q_1^2+R_3)^2}\frac{1}{q_1^2-2\mathbf{q}_1\cdot\mathbf{q}_2+q_2^2}  \nonumber \\
\times\frac{\Theta(\cos\,\theta_{11})\,\Theta(\cos\,\theta_{12})\Theta(\cos\,\theta_{21})\,\Theta(\cos\,\theta_{22})}{(\cos \theta_{11}+\cos \theta_{21})(\cos \theta_{12}+\cos \theta_{22})}\cos \theta_{12}\;\Theta(q_1-q_2 b_1) \nonumber \\
\times\left(q_1 q_2\,\cos\theta_{21}\,(1-\Theta(q_1-q_2 b_2))
+q_2^2\cos \theta_{22}\,\Theta(q_1-q_2 b_2)\frac{}{}\right)+\cdots, \quad  \label{B5}
\end{eqnarray}
where we have defined $b_1=\cos \theta_{12}/\cos \theta_{11}>0$ and $b_2=\cos \theta_{22}/\cos \theta_{21}>0$, and where the trailing dots indicate terms duplicating those explicitly given with the exception that the roles of $q_1$ and $q_2$ are interchanged. 

Limits defining the angular portions of the $\mathbf{q}$ integrations depend in a complicated way on the angle $\beta$ separating $\mathbf{k}_1$ and $\mathbf{k}_2$. In the case where the angle $\beta$ equals zero, the angular portions of the $q$-integrations in (B5) range over the $2\pi$ solid angle above a plane perpendicular to both wave vectors $\mathbf{k}_1$ and $\mathbf{k}_2$ and tangent to the unit Fermi sphere at a point an infinitesimal distance above their coincident tips. As $\beta$ increases from zero, the two wave vectors separate so that the solid angle over which the $q$ integrations are to range is reduced to $2\pi-2\beta$. This is because the half of the $\mathbf{k}_1$ tangent plane for which $x<0$ is now replaced by an upturned half plane which has been rotated counterclockwise (as viewed from a great distance out along the positive $y$-axis) through an angle $\beta$ about a fold line parallel to the $y$-axis. This guarantees that $\mathbf{k}_2$ plus an infinitesimally small $\mathbf{q}_2$ will still lie outside the unit sphere.

The extent of the azimuthal integration for $\mathbf{q}_1$ is curtailed when $\beta+\theta_{11}$ is greater than $\pi/2$. In this case the circle traced by the tip of the unit vector $\mathbf{q}_1/q_1$, with the polar angle $\theta_{11}$ fixed, intersects the above up-turned plane at the azimuthal positions $\phi_1=\pi \pm \xi_1$, where $\xi_1=\arccos(\tan(\pi/2-\beta)/\tan\,\theta_{11})$, so that the integration over the azimuthal angle of $\mathbf{q}_1$ is restricted to the range $-(\pi-\xi_1)<\phi_1<\pi-\xi_1$. A corresponding restriction applies to $\mathbf{q}_2$ where $\xi_2 =\arccos(\tan(\pi/2-\beta)/\tan\,\theta_{12})$.

An integration by parts, $vdu=uv-udv$, with $u$ the indefinite integral of either $q_1^4$ or $q_1^5$ times the reciprocal of $(q_1^2+R_3)^2(q_1^2-2q_1\,q_2\,\cos\,\psi+q_2^2)$, where $\cos\,\psi=\cos\,\theta_{11}\cos\,\theta_{12}+\sin\,\theta_{11}\sin\,\theta_{12}\,\cos(\phi_1-\phi_2)$, and with $v$ a step function, (or a product of two step functions) which vanishes at the lower limit $q_1=0$, gives a $udv$ term which includes a delta function: $\delta(q_1-q_2\,b_1)$ or $\delta(q_1-q_2\,b_2)$. The subsequent integration over $q_2$ of the explicitly written out terms in (B5) gives
\begin{eqnarray}
\frac{3\alpha R_3\log\,R_3}{4\pi^5}\left\{\int^{\pi/2}_0d\beta\sin\,\beta\left[\left(\int^{\pi/2}_{\pi/2-\beta}d\theta_{11}\sin\,\theta_{11}\int^{\pi-\xi_1}_{-(\pi-\xi_1)}d\phi_1 +\int^{\pi/2-\beta}_0 d\theta_{11}\sin\,\theta_{11}\int^{\pi}_{-\pi}d\phi_1\right)\right.\right.\nonumber \\
\times\left(\int^{\pi/2}_{\pi/2-\beta}d\theta_{12}\sin\,\theta_{12}\int^{\pi-\xi_2}_{-(\pi-\xi_2)}d\phi_2 
+\left.\left. \int^{\pi/2-\beta}_0 d\theta_{12} \sin\,\theta_{12}\int^{\pi}_{-\pi}d\phi_2\right)\right]\right. \nonumber \\
+\left.\int^{\pi}_{\pi/2}d\beta\,\sin\,\beta\int^{\pi/2}_{\beta-\pi/2}d\theta_{11} \sin\,\theta_{11}\int^{\xi_1}_{-\xi_1}d\phi_1\int^{\pi/2}_{\beta-\pi/2} d\theta_{12}\,\sin\,\theta_{12}\int^{\xi_2}_{-\xi_2}d\phi_2 \right\} \nonumber \\ \times\Theta(\cos\,\theta_{21})\;\Theta(\cos\,\theta_{22})\cos\,\theta_{12}\left[\cos\,\theta_{21}\Theta(b_2-b_1)\left(\frac{}{}I4(\psi,b_1)-I4(\psi,b_2)\right)\right.\nonumber \\
+\left.\cos\,\theta_{22}\left(\Theta(b_1-b_2)I5(\psi,b_1)+\Theta(b_2-b_1)I5(\psi,b_2)\frac{}{}\right)\right].\qquad \label{B6}
\end{eqnarray}
In (B6) the functions $I4$ and $I5$ are defined by
\begin{equation}
I4(\psi,b)=\csc\,\psi\,\arctan\,(b\csc\,\psi-\cot\,\psi) \nonumber
\end{equation}
and
\begin{equation}
I5(\psi,b)=\log \,(b^2-2b\cos\,\psi+1)/2+\cot\,\psi\arctan\,(b\csc\,\psi-\cot\,\psi)-1/2. \nonumber
\end{equation}

To evaluate the five-fold integration over angles in (B6), ten values of $\beta$, linearly distributed for $0\leq\beta\leq 0.014$, plus thirty additional values with logarithmically increasing separations over the range $0.014<\beta<\pi$, were considered. For each $\beta$ the polar angles $\theta_{11}$ and $\theta_{12}$ were each allowed to assume thirty distinct values chosen so as to implement an approximation, via a product of Chebyshev polynomials over the planar area $(0<\theta_{11}<\pi/2)\times(0<\theta_{12}<\pi/2)$. The initial double integral over $\phi_1$ and $\phi_2$ in (B6) was performed numerically for each combination of $\theta_{11}$ and $\theta_{12}$ filling a $30\times 30$ matrix for each value of $\beta$ considered. Adding each of these matrices to its transpose then accounted for the contributions of the trailing terms in (B5) for which $q_1$ and $q_2$ were interchanged. With the Chebyshev coefficients available, the double integrals over $\theta_{11}$ and $\theta_{12}$ could be evaluated with little additional computational effort. Finally cubic spline interpolation over forty $\beta$ values facilitated the final $\beta$ integration giving $0.0414\,R_3\,\log\,R_3\text{ Ry}$ for the numeric contribution of the final term of Eq.(28).

\bibliography{Main}

\begin{thebibliography}{17}
\expandafter\ifx\csname natexlab\endcsname\relax\def\natexlab#1{#1}\fi
\expandafter\ifx\csname bibnamefont\endcsname\relax
  \def\bibnamefont#1{#1}\fi
\expandafter\ifx\csname bibfnamefont\endcsname\relax
  \def\bibfnamefont#1{#1}\fi
\expandafter\ifx\csname citenamefont\endcsname\relax
  \def\citenamefont#1{#1}\fi
\expandafter\ifx\csname url\endcsname\relax
  \def\url#1{\texttt{#1}}\fi
\expandafter\ifx\csname urlprefix\endcsname\relax\def\urlprefix{URL }\fi
\providecommand{\bibinfo}[2]{#2}
\providecommand{\eprint}[2][]{\url{#2}}

\bibitem[{\citenamefont{Foulkes et~al.}(2001)\citenamefont{Foulkes, Mitas,
  Needs, and Rajagopal}}]{rev}
\bibinfo{author}{\bibfnamefont{W.~M.~C.} \bibnamefont{Foulkes}},
  \bibinfo{author}{\bibfnamefont{L.}~\bibnamefont{Mitas}},
  \bibinfo{author}{\bibfnamefont{R.~J.} \bibnamefont{Needs}}, \bibnamefont{and}
  \bibinfo{author}{\bibfnamefont{G.}~\bibnamefont{Rajagopal}},
  \bibinfo{journal}{Rev. Mod. Phys.} \textbf{\bibinfo{volume}{73}}
  (\bibinfo{year}{2001}).

\bibitem[{\citenamefont{Slater}(1960)}]{Slater}
\bibinfo{author}{\bibfnamefont{J.~C.} \bibnamefont{Slater}},
  \emph{\bibinfo{title}{Quantum Theory of Atomic Structure}},
  vol.~\bibinfo{volume}{2} (\bibinfo{publisher}{McGraw Hill},
  \bibinfo{address}{New York}, \bibinfo{year}{1960}), \bibinfo{note}{pages 1-30
  and Appendix 22}.

\bibitem[{\citenamefont{Porter}(2009)}]{P09}
\bibinfo{author}{\bibfnamefont{J.~C.} \bibnamefont{Porter}},
  \bibinfo{journal}{Phys. Rev. B} \textbf{\bibinfo{volume}{80}},
  \bibinfo{pages}{205102} (\bibinfo{year}{2009}).

\bibitem[{\citenamefont{Mayer and Mayer}(1940)}]{M}
\bibinfo{author}{\bibfnamefont{J.~E.} \bibnamefont{Mayer}} \bibnamefont{and}
  \bibinfo{author}{\bibfnamefont{M.~G.} \bibnamefont{Mayer}},
  \emph{\bibinfo{title}{Statistical Mechanics of Fluids}}
  (\bibinfo{publisher}{Wiley and Sons}, \bibinfo{address}{New York},
  \bibinfo{year}{1940}).

\bibitem[{\citenamefont{Porter}(1965{\natexlab{a}})}]{P65}
\bibinfo{author}{\bibfnamefont{J.~C.} \bibnamefont{Porter}},
  \bibinfo{journal}{Phys. Rev.} \textbf{\bibinfo{volume}{140}},
  \bibinfo{pages}{A732} (\bibinfo{year}{1965}{\natexlab{a}}).

\bibitem[{\citenamefont{Oberman et~al.}(1962)\citenamefont{Oberman, Ron, and
  Dawson}}]{ORD}
\bibinfo{author}{\bibfnamefont{C.}~\bibnamefont{Oberman}},
  \bibinfo{author}{\bibfnamefont{A.}~\bibnamefont{Ron}}, \bibnamefont{and}
  \bibinfo{author}{\bibfnamefont{J.}~\bibnamefont{Dawson}},
  \bibinfo{journal}{Phys. Fluids} \textbf{\bibinfo{volume}{5}}
  (\bibinfo{year}{1962}).

\bibitem[{\citenamefont{Fetter and Walecka}(1971)}]{FW}
\bibinfo{author}{\bibfnamefont{A.~L.} \bibnamefont{Fetter}} \bibnamefont{and}
  \bibinfo{author}{\bibfnamefont{J.~D.} \bibnamefont{Walecka}},
  \emph{\bibinfo{title}{Quantum Theory of Many-Particle Systems}}
  (\bibinfo{publisher}{McGraw-Hill}, \bibinfo{address}{New York},
  \bibinfo{year}{1971}), \bibinfo{note}{page 156}.

\bibitem[{\citenamefont{Gell-Mann and Breuckner}(1957)}]{GB}
\bibinfo{author}{\bibfnamefont{M.}~\bibnamefont{Gell-Mann}} \bibnamefont{and}
  \bibinfo{author}{\bibfnamefont{K.~A.} \bibnamefont{Breuckner}},
  \bibinfo{journal}{Phys. Rev.} \textbf{\bibinfo{volume}{106}}
  (\bibinfo{year}{1957}).

\bibitem[{\citenamefont{Carr and Maradudin}(1964)}]{Carr}
\bibinfo{author}{\bibfnamefont{J.}~\bibnamefont{Carr}, \bibfnamefont{W.~J.}}
  \bibnamefont{and} \bibinfo{author}{\bibfnamefont{A.~A.}
  \bibnamefont{Maradudin}}, \bibinfo{journal}{Phys. Rev.}
  \textbf{\bibinfo{volume}{133}} (\bibinfo{year}{1964}).

\bibitem[{\citenamefont{Farmelo}(2009)}]{Dirac}
\bibinfo{author}{\bibfnamefont{G.}~\bibnamefont{Farmelo}},
  \bibinfo{journal}{Physics Today} \textbf{\bibinfo{volume}{46}}
  (\bibinfo{year}{2009}).

\bibitem[{\citenamefont{Korn and Korn}(1968)}]{KK}
\bibinfo{author}{\bibfnamefont{G.~A.} \bibnamefont{Korn}} \bibnamefont{and}
  \bibinfo{author}{\bibfnamefont{T.~M.} \bibnamefont{Korn}},
  \emph{\bibinfo{title}{Mathematical Handbook for Scientists and Engineers
  Second Ed.}} (\bibinfo{publisher}{McGraw Hill}, \bibinfo{address}{New York},
  \bibinfo{year}{1968}).

\bibitem[{\citenamefont{Porter}(1965{\natexlab{b}})}]{thesis}
\bibinfo{author}{\bibfnamefont{J.~C.} \bibnamefont{Porter}}, Ph.D. thesis,
  \bibinfo{school}{Purdue University} (\bibinfo{year}{1965}{\natexlab{b}}).

\bibitem[{not()}]{note}
\bibinfo{note}{Eq.(2) should replace the corresponding expressions in
  references 3 and 5.}

\bibitem[{N3()}]{N3}
\bibinfo{note}{The final term of Eq.(18) was incorrectly omitted from reference
  8.}

\bibitem[{\citenamefont{Onsager et~al.}(1966)\citenamefont{Onsager, Mittag, and
  Stephen}}]{Onsager}
\bibinfo{author}{\bibfnamefont{L.}~\bibnamefont{Onsager}},
  \bibinfo{author}{\bibfnamefont{L.}~\bibnamefont{Mittag}}, \bibnamefont{and}
  \bibinfo{author}{\bibfnamefont{M.~J.} \bibnamefont{Stephen}},
  \bibinfo{journal}{Ann Physik} \textbf{\bibinfo{volume}{18}},
  \bibinfo{pages}{71} (\bibinfo{year}{1966}).

\bibitem[{\citenamefont{Rajagopal and Kimball}(1977)}]{RK}
\bibinfo{author}{\bibfnamefont{A.~K.} \bibnamefont{Rajagopal}}
  \bibnamefont{and} \bibinfo{author}{\bibfnamefont{J.~C.}
  \bibnamefont{Kimball}}, \bibinfo{journal}{Phys. Rev. B}
  \textbf{\bibinfo{volume}{15}} (\bibinfo{year}{1977}).

\bibitem[{\citenamefont{Montgomery and Tidman}(1964)}]{MT}
\bibinfo{author}{\bibfnamefont{D.~C.} \bibnamefont{Montgomery}}
  \bibnamefont{and} \bibinfo{author}{\bibfnamefont{D.~A.}
  \bibnamefont{Tidman}}, \emph{\bibinfo{title}{Plasma Kinetic Theory}}
  (\bibinfo{publisher}{McGraw Hill}, \bibinfo{address}{New York},
  \bibinfo{year}{1964}), \bibinfo{note}{see Ch.6}.

\end{thebibliography}
\end{document}